\documentclass[conference]{IEEEtran}
\IEEEoverridecommandlockouts
\usepackage{cite}
\usepackage{amsmath,amssymb,amsfonts}
\usepackage{algorithmic}
\usepackage{graphicx}
\usepackage{textcomp}
\usepackage{xcolor}
\usepackage{subcaption}
\usepackage[labelfont=bf]{caption}
\def\BibTeX{{\rm B\kern-.05em{\sc i\kern-.025em b}\kern-.08em
    T\kern-.1667em\lower.7ex\hbox{E}\kern-.125emX}}
\begin{document}

\title{Hybrid Y-Net Architecture for Singing Voice Separation \\

}

\makeatletter
\newcommand{\linebreakand}{%
  \end{@IEEEauthorhalign}
  \hfill\mbox{}\par
  \mbox{}\hfill\begin{@IEEEauthorhalign}
}
\makeatother

\author{
    \IEEEauthorblockN{ Rashen Fernando}
    \IEEEauthorblockA{\textit{dept. Electrical \& Electronic Eng.} \\
    \textit{University of Peradeniya}\\
    Peradeniya, Sri Lanka \\
    e16103@eng.pdn.ac.lk}
  \and
    \IEEEauthorblockN{ Pamudu Ranasinghe}
    \IEEEauthorblockA{\textit{dept. Electrical \& Electronic Eng.} \\
    \textit{University of Peradeniya}\\
    Peradeniya, Sri Lanka \\
    e16306@eng.pdn.ac.lk}
  \and
    \IEEEauthorblockN{ Udula Ranasinghe}
    \IEEEauthorblockA{\textit{dept. Electrical \& Electronic Eng.} \\
    \textit{University of Peradeniya}\\
    Peradeniya, Sri Lanka \\
    e16309@eng.pdn.ac.lk}
  \linebreakand 
    \IEEEauthorblockN{Janaka Wijayakulasooriya}
    \IEEEauthorblockA{\textit{dept. Electrical \& Electronic Engineering} \\
    \textit{University of Peradeniya}\\
    Peradeniya, Sri Lanka \\
    jan@eng.pdn.ac.lk}
  \and
    \IEEEauthorblockN{Pantaleon Perera}
    \IEEEauthorblockA{\textit{dept. Engineering Mathematics} \\
    \textit{University of Peradeniya}\\
    Peradeniya, Sri Lanka \\
    pperera@pdn.ac.lk}
}

\maketitle

\begin{abstract}
This research paper presents a novel deep learning-based neural network architecture, named Y-Net, for achieving music source separation. The proposed architecture performs end-to-end hybrid source separation by extracting features from both spectrogram and waveform domains. Inspired by the U-Net architecture, Y-Net predicts a spectrogram mask to separate vocal sources from a mixture signal. Our results demonstrate the effectiveness of the proposed architecture for music source separation with fewer parameters. Overall, our work presents a promising approach for improving the accuracy and efficiency of music source separation.

\end{abstract}

\begin{IEEEkeywords}
    Short-Time Fourier Transform, spectrogram, waveform, U-Net, source separation, Y-Net
\end{IEEEkeywords}

\section{Introduction}

Music source separation is a complex task that aims to separate individual sources in an audio mixture, which has practical applications in creating karaoke tracks, transcribing music, and music production. Deep neural networks (DNNs) have been used to effectively perform music source separation, with two main approaches utilizing time-frequency or time-domain representations. In this paper, the singing voice was separated from an audio mixture using DNN.

Although the time-frequency contains many patterns to be discovered, the Short Time Fourier Transform (STFT) is not necessarily optimal for music source separation \cite{b20}. 
The vocal data is typically present in the low-frequency bands, which demands higher resolution in these regions. However, there exists a trade-off between the time and the frequency resolution when utilizing spectrogram representation. Feeding the magnitude spectrogram into the network has been the common approach. This methodology can result in an incomplete representation of the audio signals as it lacks phase information. A recent publication titled ‘Investigating U-Nets with various intermediate blocks for spectrogram-based singing voice separation' \cite{b1} proposed a method to combine the phase spectrogram and magnitude spectrogram before feeding them into the network, thus providing additional information to enhance the separation performance. However, the weight distribution analysis of the network suggested that the phase information was not heavily utilised in the separation task.


Prior research has investigated the feasibility of incorporating all the information available in the raw audio signal, including the phase, by utilizing end-to-end models for music source separation. This approach aims to prevent the exclusion of potentially valuable information \cite{b2}, \cite{b3}, \cite{b4}. Even though raw audio signals contain both magnitude and phase information, it can be difficult to learn the underlying structure of such signals from the raw audio alone. It is evident that different representations can lead to different artefacts.

In this paper, we propose a novel architecture that combines the strengths of raw audio and spectrogram representations to estimate a time-frequency mask for separating singing voice. The proposed architecture consists of two main components: a learnable filter and a Y-net. The learnable filter offers the advantage of learning filters that are capable of extracting crucial features which are not present in the STFT spectrogram while incorporating phase information. The Y-net extracts features from both domains and recreate the spectrogram of the singing voice.

\section{Related Work}

Our network is inspired by the U-Net architecture \cite{b5}. It uses a stack of convolutional layers to encode the image into a small and deep representation, merges at the bottom layer, and adds additional skip connections between layers at the same hierarchical level in the encoder and decoder to allow low-level information to flow directly from the high-resolution input to the high-resolution output.
In the field of music source separation, the U-Net architecture is widely employed for the estimation of a mask that can separate a single instrument from a mixture signal spectrogram \cite{b6}. In contrast, spectrogram-channels U-Net \cite{b7} directly generates the spectrogram instead of estimating a mask. This approach allows the separation of multiple instruments, but it may produce unwanted artefacts in the audio signal.

The proposed method leverages raw audio and spectrogram data to extract the best possible features from the mixed audio, with the aim of separating the singing voice by generating a mask in the time-frequency domain. In this study, a novel technique was introduced to convert raw audio data into a spectrogram dimension. Specifically, the learnable spectrogram was designed to learn filter coefficients, which are then used to extract relevant features from the mixture audio waveform. This approach draws inspiration from the wavelet transform methodology \cite{b8}. The learnable spectrogram is based on Spectnet \cite{b9}, which employs learnable 1-D convolutional filters to extract distinct frequencies from the input waveform and create a spectral image.

Waveform domain source separation was first explored by 'End-to-end music source separation' \cite{b2}, that employed dilated convolutions but failed compared to spectrogram-driven models at the time. Convtasnet \cite{b3}, a model based on masking over a learnt time-frequency representation using dilated convolutions, outperformed the existing state-of-the-art models based on the spectral domain in 2019, as end-to-end waveform models incorporate phase information. But it suffered from significant audio artefacts as measured by human evaluations. An improved novel waveform-to-waveform model with a U-Net structure and bidirectional LSTM was introduced in 2021 \cite{b10}, which was built upon Wave-U-Net \cite{b4}, an adaptation of the U-Net to the one-dimensional time domain that outperformed Convtasnet. However, Wave-U-Net uses faster strided convolutions with a larger number of channels, which can introduce aliasing artefacts.

E.M. Grais \cite{b11} highlights the importance of time-frequency resolution in spectrograms for feature extraction, where low and high-frequency bands are processed with different-sized kernels. These papers question the reliability of the spectrogram.
Learnable spectrograms offer improved reliability in capturing low frequencies at higher resolutions compared to traditional spectrograms, as weight biases have shown to be effective in this regard. This represents a valuable advancement in music source separation techniques.
J. Paulus and M. Torcoli \cite{b12} shows that sampling input mixture audio, either in 8kHz or 44kHz does not improve the separation quality of vocals significantly as neural networks manage to extract the same features in the U-Net core. Though speech signals are present at low frequencies (8 kHz), singing voices can occasionally reach up to 17 kHz. A spectrogram that is trainable can determine the optimal features to extract from the raw audio for isolating the vocals, while a Y-net can recover any missing features from the input spectrogram. Hybrid spectrogram and waveform source separation \cite{b13} extends the original U-Net architecture and provides two parallel branches: one in the time (temporal) domain and one in the frequency (spectral) domain, with shared features in the U-Net core while having a decoder for each branch. Hybrid Transformer Demucs \cite{b14} replaces the innermost layers of the two U-Net architectures with transformer layers, which then requires large amounts of data for training.

In contrast to our proposed Y-shaped architecture, previous studies have adopted an X-shaped architecture where separate branches were used for each input \cite{b14}, \cite{b13}. Our approach utilises a single decoder branch, resulting in lower parameter count, faster training, lower computational cost, and less data requirement than the aforementioned studies.

\begin{figure}[h]
\centerline{\includegraphics[width=85mm,scale = 0.5]{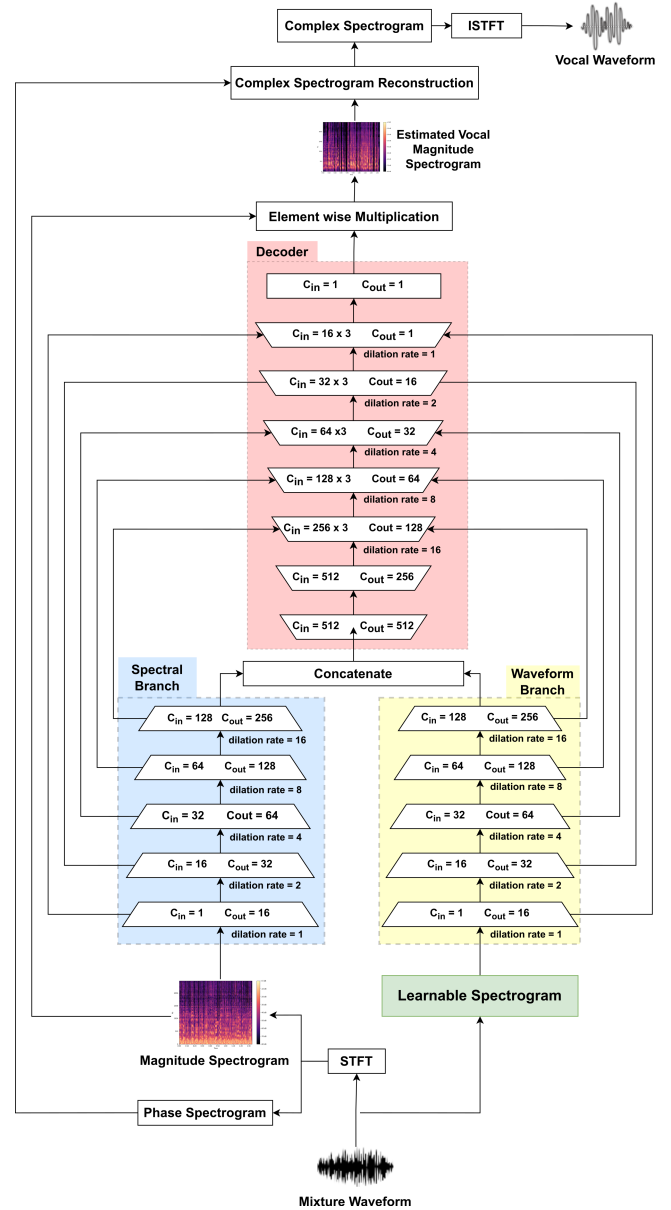}}
\caption{ Y-Net Architecture : Takes the raw audio waveform and the magnitude spectrogram as inputs, and outputs the magnitude spectrogram of the separated vocal}
\label{fig:Fig 1}
\end{figure}

\section{Architecture}
This paper presents a neural network architecture that takes raw audio waveform and spectrogram as inputs, and estimates a mask similar to the shape of input spectrogram that can be used to separate and extract singing voice from a mixture. Specifically, the estimated mask is multiplied with the original mixture spectrogram to obtain the magnitude spectrogram of the singing voice.

It is worth noting that our network does not alter the phase of the original mixture signal. This approach is similar to the convolutional recurrent neural network with the attention framework proposed by C. Sun et al. for speech separation in monaural recordings \cite{b15}. To obtain the final predicted source signals, we combine the estimated magnitude spectrogram with the phase spectrogram of the original mixture signal to create the complex spectrogram. We apply the Inverse Short-Time Fourier Transform (ISTFT) to obtain the vocal waveform as shown in Figure 1. By preserving the phase information, the predicted source signals retain the original temporal structure and sound coherence.
 
Our network consists of two encoder branches, namely the waveform branch and the spectral branch, which take the raw audio and spectrogram, respectively. The front end of the raw audio encoder branch includes a learnable spectrogram module, which converts the raw audio to a spectrogram format. Then two encoders extract features from both inputs that are then merged at the network core and sent to the decoder. The decoder consists of a stack of up-sampling layers which decodes the features to the original size of the spectrogram, as shown in Figure \ref{fig:Fig 1}. It is worth noting that the two encoders and the decoder structure are symmetric.

\begin{figure}[h]
    \centerline{\includegraphics[width=88mm,scale = 0.5]{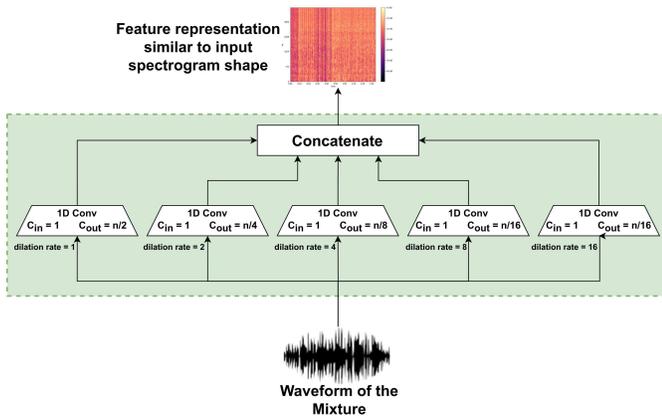}}
    \caption{The model takes the raw audio waveform as input and extracts features using 1-D convolution layers. The output is a feature representation that is similar in size to an input spectrogram.}
    \label{fig: Fig 2}
\end{figure}

\subsection{Learnable spectrogram}
The architecture of the learnable spectrogram is depicted in Figure \ref{fig: Fig 2}. In order to create the learnable spectrogram, initially, a single channel of audio is convoluted with five sets of 1-D convolutions with varying dilation rates (1, 2, 4, 8, 16), which allows the network to capture long-term dependencies. For each set, convolution layers with a lower dilation rate get a higher number of kernels dedicated because local features affect the vocals more dominantly than global features. Therefore, we give a higher priority to the set of convolutions with dilation rate 1. The sum of all the kernels in five sets of convolution layers equals the number of frequency bins in the input spectrogram, which is 1024 in our model.
 
The kernel of length 2048 is strided with a varying overlap between consecutive frames, forming vectors of length equal to the number of time bins in the input spectrogram. Finally, all the convoluted signals are concatenated together, forming an output with a shape similar to the input spectrogram. LeakyRelu is used as the activation function after each 1-D convolution. Here, kernels act as filters similar to the mother wavelets in the wavelet transform.

\begin{figure}[h]
  \centering 
    \begin{subfigure}[t]{0.25\textwidth}
        \centering
        \includegraphics [width=43mm, scale = 0.8]{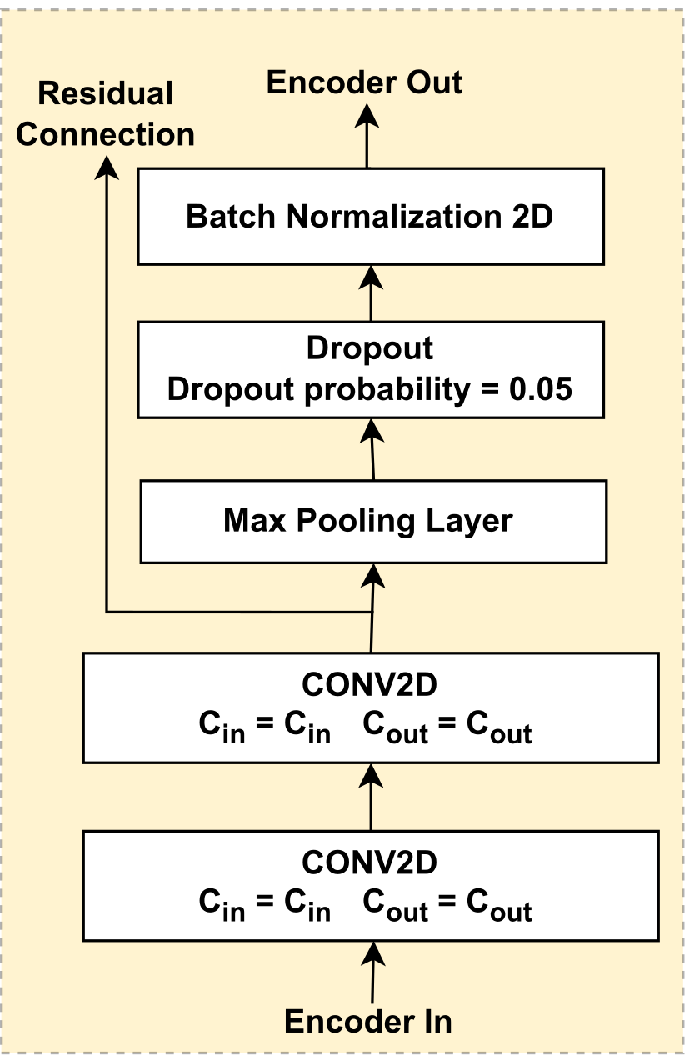}
        \caption{Encoder Layer}
        \label{fig:encoder}
    \end{subfigure}%
    \begin{subfigure}[t]{0.25\textwidth}
        \centering
        \includegraphics[width=43mm]{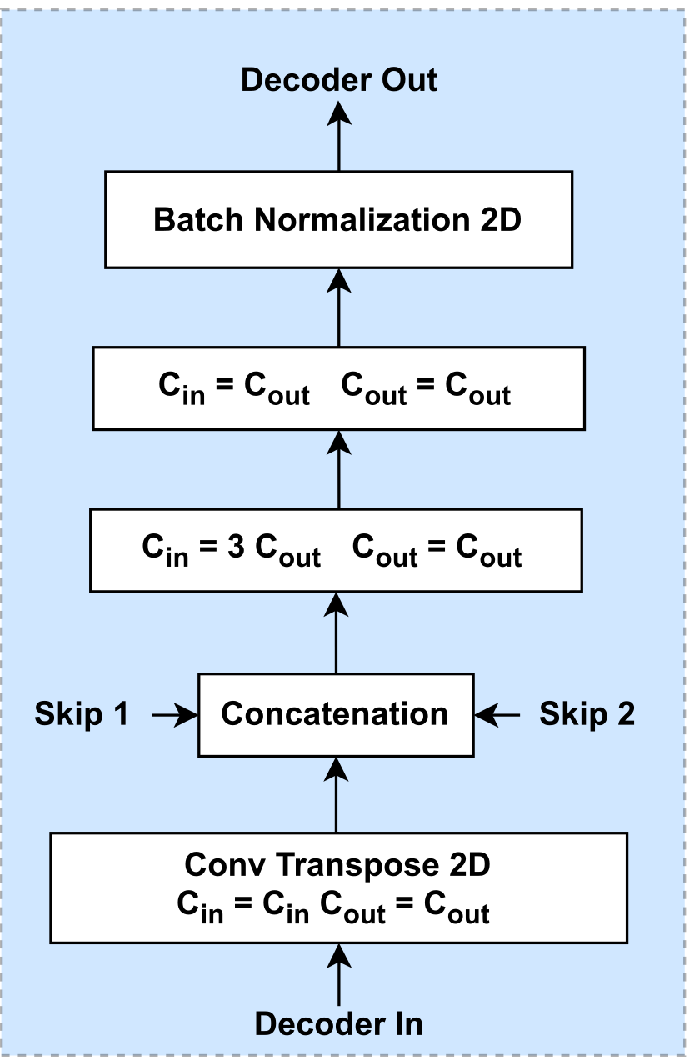}
        \caption{Decoder Layer}
        \label{fig:decoder}
    \end{subfigure}
    
    \captionsetup{labelfont=bf}
    \caption{(a): The Encoder layer with 2D convolution layers and a max pooling layer with a skip connection going out before the max pooling layer. (b): The Decoder layer with an up-sampling layer followed by two 2D convolution layers with two skip connections coming from the encoder layers in spectral and waveform branches that get concatenated before the convolution layer  }
    \label{fig:Fig 3}
\end{figure}

\subsection{Encoder}
The representation from the learnable spectrogram is then sent through five encoder layers that double the number of channels with increasing dilation rate while halving the size of the input spectral image as it goes deep. Each encoder layer consists of a batch normalization layer followed by two convolution layers with the kernel size of 3x3, a max pooling layer, and a dropout layer. LeakyRelu is used as the activation function after each convolution. The architecture of a single encoder layer is depicted in Figure \ref{fig:Fig 3} (\subref{fig:encoder}).

Before feeding the spectral branch, we transform the raw audio mixture waveform to a spectrogram using STFT. In our experiment, the input spectrogram consisted of 1024 frequency bins and 128 time bins. The spectrogram is sent through a similar five-encoder structure to the waveform branch, except the activation function after each convolution in the encoder is replaced by ReLu. Following the fifth encoder block, the network concatenates features that have been extracted from both the spectral and waveform branches, provided they are of equal size. This concatenation takes place at the network's core and is subsequently followed by two convolutional layers and a batch normalization layer. These concatenated features are then activated using ReLu before being forwarded to the decoder component of the network.

\subsection{Skip connections}
Skip connections are an important aspect of our architecture because they propagate the information from both encoder branches to a shared decoder branch, bypassing the intermediate processing stages within the network. The use of skip connections helps to mitigate the effects of down-sampling and up-sampling operations, which can otherwise cause important information in the audio signal to be lost. Additionally, skip connections provide a direct connection between the encoder and decoder layers of the network, allowing the model to better preserve high-frequency details and fine-grained structures in the output audio signal.

\subsection{Decoder}

Encoder output is sent through 5 decoder layers where each decoder layer consists of an up-sampling layer followed by two convolution layers activated with the ReLu function with a kernel size of 3x3 and a batch normalization layer. For each decoder layer, there are three inputs coming from the previous decoder layer and two skip connections coming from the encoder layers in both the spectral and waveform branches. The final decoder layer output is then sent through a convolution layer of kernel size 1x1, which is then activated by Hardtanh in order to enhance the vocal features and suppress other features in the frequency domain. The architecture of a single decoder layer is depicted on Figure \ref{fig:Fig 3} (\subref{fig:decoder}).
 
Finally, the decoder output gives the estimated mask. Element-wise multiplication of the decoder output with the input mixture spectrogram gives the spectrogram of the singing voice.
 
In summary, the encoder-decoder architecture provides a way to extract meaningful features from both inputs and then use these features to generate a mask to separate the singing voice.

\section{Results and Experiment}
\subsection{Dataset}

MUSDB18 is a dataset that is commonly used for research in the field of music source separation \cite{b21}. 
It is a compilation of professionally produced music tracks that spans various genres and styles. These tracks consist of multiple sources, such as vocals, drums, bass, and other instruments. We used a total of 150 songs from the MUSDB18 dataset, with 100 songs for training, 25 songs for validation, and 25 songs for testing. Initially, we take the mono vocal, and the mixture audio tracks sampled at 44 kHz and slice them into 67072 sample time windows, which correspond to approximately 1.5 seconds. 

\subsection{Experimental setup}

We developed the model in the Pytorch framework. The neural network was trained for 100 epochs over the MUSDB18 dataset, using an Nvidia V100 GPU with 32GB of memory. During each epoch, the entire dataset was passed through the network in batches of size 16, resulting in a total of 497 batches. The mixture audio waveform and the STFT magnitude spectrogram were used as the training inputs, while the corresponding vocal spectrogram was the ground truth for the training process.
A single data element of the waveform domain contains 67072 data points. It is windowed at 2048 samples with 75\% overlap to obtain the magnitude spectrogram, resulting in 128 time bins and 1024 frequency bins. As the loss function, we used the Mean Squared Error (MSE), where the training objective was to minimize the MSE between the ground truth spectrogram and the estimated spectrogram. We used the Adam optimizer \cite{b16} with a learning rate of 0.0001.


\subsection{Evaluation metric}

To assess the performance of source separation objectively, we employ the Signal-to-Distortion Ratio (SDR) \cite{b17}, Scale-Invariant Signal to-Noise-Ratio (SI-SNR) \cite{b3}, and Short-Time Objective Intelligibility (STOI) \cite{b18} as evaluation metrics for our models. A higher value for SDR, SI-SNR, and STOI indicates better quality. While SDR and SI-SNR values can exceed 1, STOI values fall within the range of 0 to 1. Table \ref{tab:Table 1} presents an evaluation of the network modifications, namely U-Net, U-Net + learnable spectrogram, and Y-Net, using the above-mentioned metrics. 

\subsection{U-Net}

Initially, we utilised the U-Net as our baseline architecture for singing voice separation where the input is a mixture audio spectrogram. However, we encountered challenges with the performance of conventional spectrograms, which we have discussed in the previous sections of our paper. As a result, we explored alternative strategies. The U-Net generates a mask at its output, which can be multiplied with the input mixture spectrogram to derive the singing voice spectrogram. While the objective evaluation metrics exhibit decent performance on the mask estimation method, we observed that estimating the vocal spectrogram directly suppresses the accompaniment parts that are being leaked into the vocal spectrogram effectively, leading to improved subjective performance but a loss in SDR. In order to compensate for the lack of human subjects in our study, we have chosen to focus exclusively on objective metrics.

\subsection{U-Net + Learnable spectrogram}

Our proposition was to add a learnable spectrogram as the front end for the U-Net architecture that transforms raw audio into a format similar to a spectrogram, which contains features that are suitable for source separation. We integrated this approach with a U-Net architecture to input, mixture waveform and predict the vocal spectrogram directly at the U-Net end. However, the performance of this method was inferior to that of the U-Net with a spectrogram input. Specifically, using learnable spectrogram techniques caused a decline of 1.86 dB in SDR. On the other hand, we observed that this method requires a larger dataset because understanding the underlying structure of the audio is challenging.


\begin{figure}[h]
  \centering 
    \begin{subfigure}[b]{0.16\textwidth}
        \centering
        \includegraphics [width=29mm]{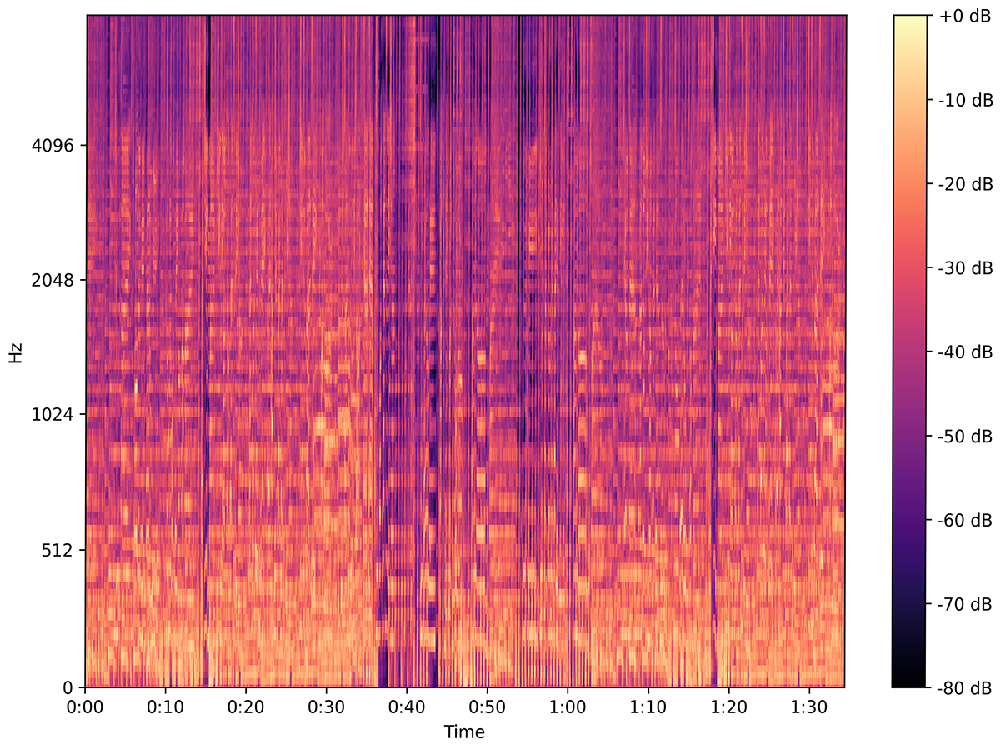}
        \caption{ }
        \label{fig:mixture}
    \end{subfigure}%
    \begin{subfigure}[b]{0.16\textwidth}
        \centering
        \includegraphics[width=29mm]{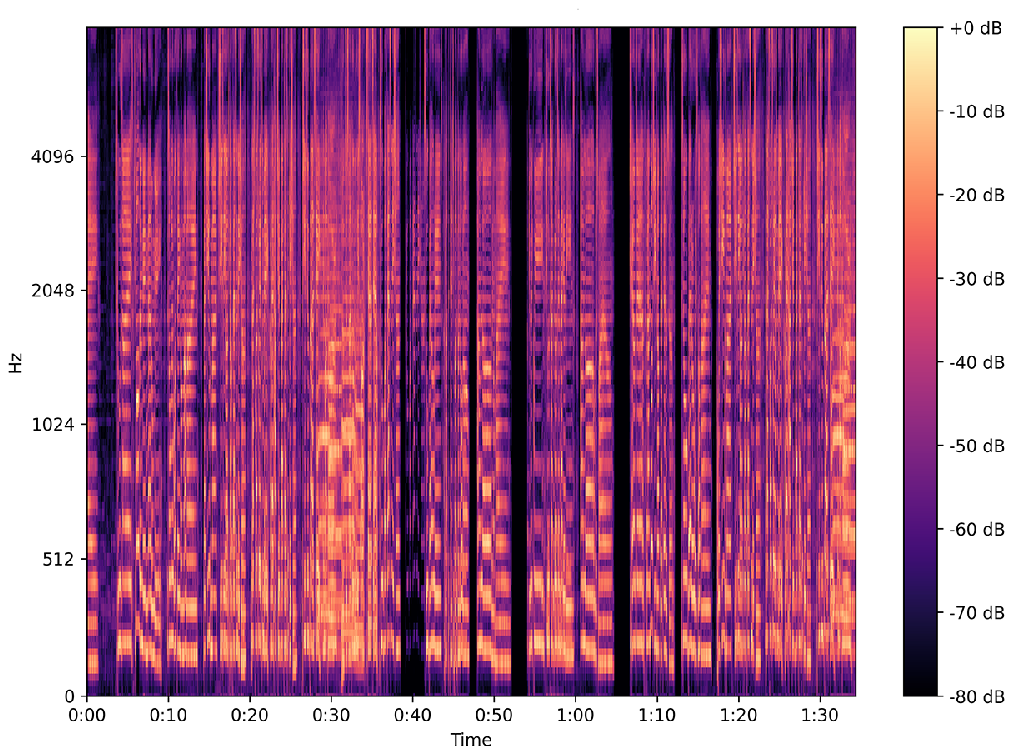}
        \caption{ }
        \label{fig:Ground truth}
    \end{subfigure}%
    \begin{subfigure}[b]{0.16\textwidth}
        \centering
        \includegraphics[width=29mm]{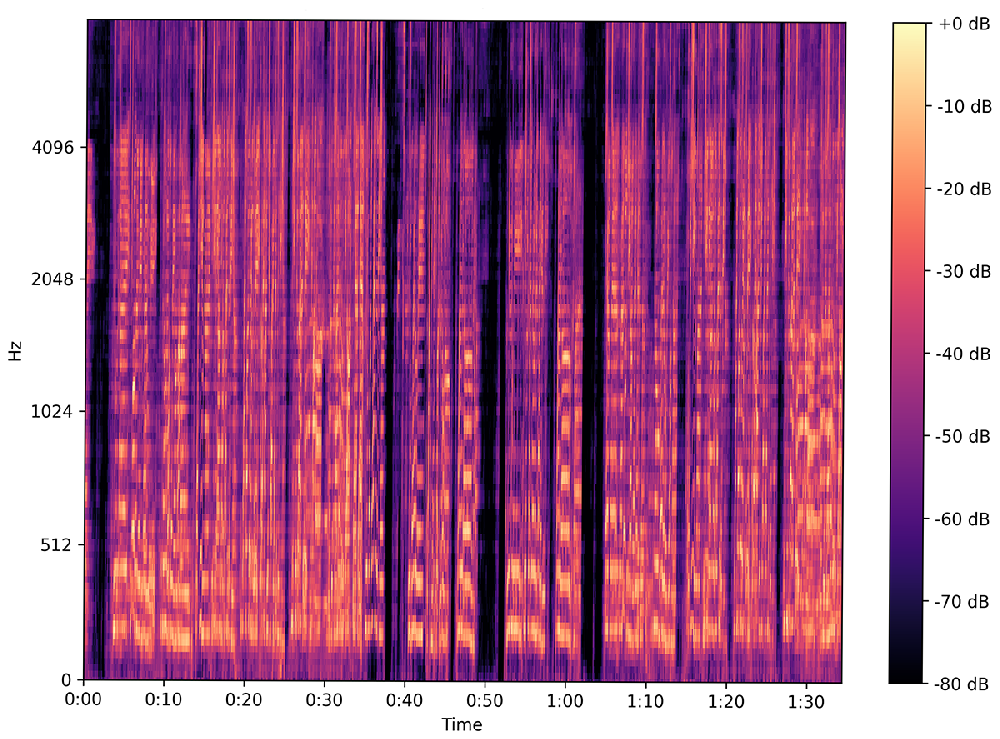}
        \caption{ }
        \label{fig:Estimated}
    \end{subfigure}
    \caption{The Mel spectrogram of the (a): Mixture audio signal  (b): Singing voice (ground truth) (c): Singing voice estimated by the Y-Net }
    \label{fig:Fig 4}
\end{figure}

\subsection{Y-Net}

We proceeded to investigate a Y-shaped network that concatenated the features extracted from both the spectral and waveform domains. This approach substantially enhanced the performance of the source separation task compared to our baseline, resulting in a 0.95 dB increase in SDR. Additionally, we enhanced the network by utilizing a leaky ReLu activation function for the waveform branch while the ReLu function was applied in the spectral branch. We experimented with various activation functions for the final layer, such as threshold, ReLu, sigmoid, and Hardtanh. Through empirical analysis and experimentation, we discovered that integrating a Hardtanh activation function to the final layer led to a significant improvement in performance, with less training compared to before. Figure \ref{fig:Fig 4} shows the Mel-spectrograms of the mixture, ground truth (vocal) and the results obtained from the Y-Net (separated vocal).


\begin{table}[h]
\caption{Comparison of Source Separation Performance of the Vocal Source Between the U-Net Model, Learnable Spectrogram + U-Net and the Y-Net. Higher SDR values are better }

\begin{center}
\label{tab:Table 1}
 \begin{tabular}{|l|l|l|l|}
    \hline
    \textbf{Method}  &\textbf{SDR} & \textbf{SI-SNR} & \textbf{STOI} \\
    \hline
    U-Net & 6.61 & 5.82 & 0.55 \\
    U-Net+learnable spectrogram & 4.75 & 3.93 & 0.54 \\
    Y-Net & \textbf{7.56} & \textbf{6.13} & \textbf{0.62} \\
    
    \hline
  \end{tabular}
  
  \end{center}
\end{table}

\section{Conclusion}
The goal of this research was to improve source separation for music in the MUSDB18 dataset by modifying the U-Net architecture into a Y-Net. Our modifications include incorporating parallel waveform and spectrogram encoder branches that converge into a common core, which allows the leveraging of both spectrogram and raw audio features through a learnable spectrogram-based raw audio feature extraction method. However, the reduction of SDR introduced by the ISTFT, while obtaining the waveform from an estimated spectrogram is a limitation of our method. A potential solution to this issue might be the use of an end-to-end waveform network. 




\begin{thebibliography}{00}

\bibitem{b20} M. Goto and H. Hashiguchi, "Improved harmonic-percussive separation based on masking principle and its application to real-world audio signals," in Proceedings of the IEEE International Conference on Acoustics, Speech, and Signal Processing (ICASSP), 2003, pp. 105-108.



\bibitem{b1} W. Choi, M. Kim, J. Chung, D. Lee, and S. Jung, ‘Investigating u-nets with various intermediate blocks for spectrogram-based singing voice separation’, ISMIR 2020.



\bibitem{b2}	F. Lluís, J. Pons, and X. Serra, ‘End-to-end music source separation: is it possible in the waveform domain?’, in Interspeech, 2018.



\bibitem{b3} Y. Luo and N. Mesgarani, ‘Conv-tasnet: Surpassing ideal time--frequency magnitude masking for speech separation’, IEEE/ACM transactions on audio, speech, and language processing, vol. 27, no. 8, pp. 1256–1266, 2019.


\bibitem{b4} D. Stoller, S. Ewert, and S. Dixon, ‘Wave-u-net: A multi-scale neural network for end-to-end audio source separation’, ismir2018.
 

\bibitem{b5} A. Jansson, E. Humphrey, N. Montecchio, R. Bittner, A. Kumar, and T. Weyde, ‘Singing voice separation with deep u-net convolutional networks’, 2017.

\bibitem{b6} V. S. Kadandale, J. F. Montesinos, G. Haro, and E. Gómez, ‘Multi-channel u-net for music source separation’, in 2020 IEEE 22nd international workshop on multimedia signal processing (MMSP), 2020, pp. 1–6.

\bibitem{b7}  J. Oh, D. Kim, and S.-Y. Yun, ‘Spectrogram-channels u-net: a source separation model viewing each channel as the spectrogram of each source’, arXiv preprint arXiv:1810. 11520, 2018.

\bibitem{b8} P. M. Bentley and J. T. E. McDonnell, ‘Wavelet transforms: an introduction’, Electronics \& communication engineering journal, vol. 6, no. 4, pp. 175–186, 1994.

\bibitem{b9} M. Ansari, T. Hasan, and Others, ‘SpectNet: End-to-End Audio Signal Classification Using Learnable Spectrograms’, arXiv preprint arXiv:2211. 09352, 2022.

\bibitem{b10} A. Défossez, N. Usunier, L. Bottou, and F. Bach, ‘Music source separation in the waveform domain’, ICLR 2020.

\bibitem{b11} E. M. Grais, F. Zhao, and M. D. Plumbley, ‘Multi-band multi-resolution fully convolutional neural networks for singing voice separation’, in 2020 28th European Signal Processing Conference (EUSIPCO), 2021, pp. 261–265.

\bibitem{b12} J. Paulus and M. Torcoli, ‘Sampling Frequency Independent Dialogue Separation’, in 2022 30th European Signal Processing Conference (EUSIPCO), 2022, pp. 160–164.

\bibitem{b13} A. Défossez, ‘Hybrid spectrogram and waveform source separation’, arXiv preprint arXiv:2111. 03600, 2021.

\bibitem{b14} S. Rouard, F. Massa, and A. Défossez, ‘Hybrid Transformers for Music Source Separation’, arXiv preprint arXiv:2211. 08553, 2022.

\bibitem{b15}  C. Sun et al., ‘A convolutional recurrent neural network with attention framework for speech separation in monaural recordings’, Scientific Reports, vol. 11, no. 1, p. 1434, 2021.

\bibitem{b21} Z. Rafii, A. Liutkus, F.-R. Stöter, S. I. Mimilakis, and R. Bittner, ‘The MUSDB18 corpus for music separation’. Dec-2017.

\bibitem{b16} Diederik P. Kingma and Jimmy Ba, “Adam: A method for stochastic optimization,” 2014

\bibitem{b17} Emmanuel Vincent, Remi Gribonval, and C ´ edric ´ Fevotte, “Performance measurement in blind audio ´ source separation,” IEEE transactions on audio, speech, and language processing, vol. 14, no. 4, pp. 1462–1469, 2006.

\bibitem{b18}  M. Kolbæk, Z.-H. Tan, and J. Jensen, ‘Monaural speech enhancement using deep neural networks by maximizing a short-time objective intelligibility measure’, in 2018 IEEE International Conference on Acoustics, Speech and Signal Processing (ICASSP), 2018, pp. 5059–5063.



\end{thebibliography}
\end{document}